\input{aipcheck}

\documentclass[
    ,final            
  ]
  {aipproc}

\layoutstyle{8x11double}

\newcommand{\Eq}[1]{Eq.~(\ref{#1})}
\newcommand{\Eqs}[2]{Eqs.(\ref{#1},\ref{#2})}

\newcommand{\ur}[1]{(\ref{#1})}

\newcommand{\beq}{\begin{equation}}
\newcommand{\eeq}{\end{equation}}

\newcommand{\la}[1]{\label{#1}}
\newcommand{\bea}{\begin{eqnarray}}
\newcommand{\eea}{\end{eqnarray}}
\newcommand{\ba}{\begin{array}}
\newcommand{\ea}{\end{array}}

\newcommand{\half}{{\textstyle{\frac{1}{2}}}}

\newcommand{\nn}{\nonumber}
\newcommand{\n}{\nonumber}
\newcommand{\Tr}{{\rm Tr\,}}

\begin{document}

\title{Confinement and deconfinement for any gauge group \\from dyons viewpoint}



\author{Dmitri Diakonov}{
  address={Petersburg Nuclear Physics Institute, Gatchina 188300, St. Petersburg, Russia},
altaddress={Ruhr-Universit\"at Bochum, Bochum 44780, Germany}
}

\author{Victor Petrov}{
  address={Petersburg Nuclear Physics Institute, Gatchina 188300, St. Petersburg, Russia},
altaddress={Ruhr-Universit\"at Bochum, Bochum 44780, Germany}
}

\begin{abstract}
Basing on a semiclassical picture of dyons, we present a nonperturbative model of a pure Yang--Mills
theory at any temperatures, for an arbitrary simple gauge group. We argue that at low temperatures
dyons drive the Yang--Mills system for all groups to a phase where the `eigenphases' of the Polyakov
line are, as a vector, proportional to the Weyl vector being the half sum of positive roots.
For most gauge groups it means confinement, in particular for `quarks' in any $N$-ality nonzero
representation of the $SU(N)$ gauge group. At a critical temperature there is a 1$^{\rm st}$ order
phase transition for all groups (except SU(2) where the transition is 2nd order), characterized
by a jump of Polyakov lines, irrespectively of whether the gauge group has a nontrivial center, or not.
\end{abstract}

\maketitle


\section{Effective action for the Polyakov line}

We consider the pure Yang--Mills (YM) theory based on a simple gauge group
at nonzero temperatures. The limit of zero temperature will also be considered;
in this case a nonzero temperature serves as an infrared regulator of the theory.

At any temperature $T$ except strictly zero, one can introduce the Polyakov loop variable
$L({\bf x})$ $= {\cal P}\exp\left(i\int_0^{1/T}dt\;A_4({\bf x,t})\right)$. Gauge freedom allows one
to consider the Euclidean time component matrix $A_4({\bf x})$ to be time-independent, moreover,
diagonal, meaning that it can be decomposed in Cartan generators $H_m$ where $m=1,\ldots,r$,
$\;r$ is the rank of the gauge group:
\bea\la{A4}
A_4({\bf x})&=&2\pi T\,\phi_m({\bf x}) H_m\equiv 2\pi T\,(\mbox{\boldmath$\phi$}({\bf x})\cdot {\bf H}),\\
L({\bf x})&=& \exp\left(2\pi i (\mbox{\boldmath$\phi$}({\bf x})\cdot {\bf H})\right),
\la{Pol1}\eea
where the factor $2\pi T$ has been introduced for convenience to make the $r$-dimensional
vector $\mbox{\boldmath$\phi$}$ dimensionless. Polyakov line can be considered in any representation;
the Cartan generators $H_m$ are then taken in the corresponding representation.
The static Abelian field $\mbox{\boldmath$\phi$}({\bf x})$, together with the periodic in time spatial
components of the YM field $A_{i}({\bf x},t)$ are the degrees of freedom of the theory.

Alternatively, one can use the Hamiltonian gauge $A_4=0$; then one has to integrate over
$A_{i}({\bf x},t)$ with quasi-periodic boundary condition being a gauge transformation:
$A_{i}({\bf x},\,t\!\!=\!\!1/T)=A_i({\bf x},\,t\!\!=\!\!0)^{L({\bf x})}\equiv
L^{-1}({\bf x})\left(A_i({\bf x})+i\partial_i\right)L({\bf x})$.
The full partition function is then a path integral over all gauge transformations at the
boundary~\cite{Gross:1980br}:
\bea\nn
{\cal Z}\!\!&=&\!\!\int\!DL({\bf x})\!\!\int DA_i({\bf x})\int_{A_i({\bf x})}^{A_i({\bf x})^{L({\bf x})}}\!DA_i({\bf x},t)\times\\
\la{eff-act}
\!\!&\times &\exp(-S[A_i({\bf x},t)])\;=\;\int\!D\mbox{\boldmath$\phi$}({\bf x})\times \\
\!\!\!\!\!\!&\times &\!\!\!\exp\left\{-\!\!\int\!d^3{\bf x}\left[P(\mbox{\boldmath$\phi$})\!
+\!(\partial_i\mbox{\boldmath$\phi$}\cdot\partial_i\mbox{\boldmath$\phi$})T(\mbox{\boldmath$\phi$})\!+\!\ldots\right]\right\}.
\nn\eea
The exponent in the last line is the definition of the effective action $S_{\rm eff}[\mbox{\boldmath$\phi$}]$
for the Polyakov line. A spatial gauge rotation of the Polyakov line $L({\bf x})$
can be absorbed into the $A_i({\bf x})$ of the initial condition over which one also has to integrate. Therefore,
the effective action  depends only on the gauge-invariant eigenvalues of the Polyakov line, that are related
to the `eigenphases' $\mbox{\boldmath$\phi$}$ according to \Eq{Pol1}. The derivative expansion goes in
short derivatives, and the functions $P(\mbox{\boldmath$\phi$}),T(\mbox{\boldmath$\phi$}),...$ are periodic~\cite{Diakonov:2004kc}.

In perturbation theory, the first two terms of the effective action are well known~\cite{Gross:1980br,Diakonov:2004kc,Weiss:1980rj};
in the one-loop approximation the potential energy, as function of the phases $\mbox{\boldmath$\phi$}$, is
\beq
P^{\rm pert}(\mbox{\boldmath$\phi$})=\frac{2\pi^2}{3}T^3\sum_{\alpha}\!(\mbox{\boldmath$\phi$}\cdot\mbox{\boldmath$\alpha$})^2
\left(\!1\!-|(\mbox{\boldmath$\phi$}\cdot\mbox{\boldmath$\alpha$})|\right)^2
\la{P-pert}\eeq
where the sum goes over all root vectors $\mbox{\boldmath$\alpha$}$ of the gauge group.
The quantities $2\pi T (\mbox{\boldmath$\phi$}\cdot\mbox{\boldmath$\alpha$})$ are the `charged gluon' masses
in the background of a constant $A_4$. More precisely, they are the eigenvalues of $A_4$ in the adjoint
representation.

The potential has the minimum at $\mbox{\boldmath$\phi$}=0$, and at its periodic repeats.
If the gauge group has a non-trivial center, as {\it e.g.} in $SU(N)$, the potential energy \ur{P-pert} has
symmetric minima at all values of $L$ belonging to the center. At high temperatures, it forces
the Polyakov line to have small oscillations about one of the elements of the center.

An intriguing question (which could be answered by direct lattice simulation of the partition function
in the form of \Eq{eff-act}) is what is the potential energy in a general, nonperturbative case of
arbitrary temperatures. We argue that in the confinement phase for any gauge group the nonperturbative
potential energy has a minimum at the universal value of $\mbox{\boldmath$\phi$}_{\rm min}={\bf v}$
proportional to the so-called Weyl vector $\mbox{\boldmath$\rho$}=\half\sum_{\alpha>0}\mbox{\boldmath$\alpha$}$,
the half-sum of positive roots:
\beq
{\bf v}={\bf\rho}\;\frac{2}{c_2\mbox{\boldmath$\alpha$}_{\rm max}^2}
\la{v-opt}\eeq
where $c_2$ is the dual Coxeter number and $\mbox{\boldmath$\alpha$}_{\rm max}$ is the length square
of the longest root of the group. For all simply-laced groups of the A,D,E series
$\mbox{\boldmath$\alpha$}_{\rm max}^2=2$~\footnote{Throughout the paper we use the normalization
of the root and weight vectors from the book~\cite{Bourbaki}.
To compare different groups, we introduce the gauge coupling constant as $\exp(-\int \Tr_{\rm adj}F_{\mu\nu}^2/(4g^2c_2))$
where $c_2$ is the so-called dual Coxeter number, see below. For $SU(N)$, $c_2=N$. The one-loop running of the coupling
constant is then given by $2\pi/\alpha_s(\mu)=8\pi^2/g^2(\mu)=(11/3)c_2\ln(\mu/\Lambda)$.}.

The trace of the Polyakov loop \ur{Pol1} computed at this optimal value of $A_4$ is, for an arbitrary representation
$R$ labeled by the eldest weight ${\bf W}$,
\bea\la{Pol2}
\Tr_R\,L&=&\sum_{{\rm weights}\;{\bf w}\in R}\exp\left(2\pi i ({\bf w}\cdot {\bf v})\right)\\
\nn
&=&\prod_{\alpha>0}\frac{\sin\left(\frac{2\pi}{c_2\alpha_{\rm max}^2}({\bf W}+\rho\cdot \mbox{\boldmath$\alpha$})\right)}
{\sin\left(\frac{2\pi}{c_2\alpha_{\rm max}^2}(\rho\cdot \mbox{\boldmath$\alpha$})\right)}\,.
\eea
Here the first line is the definition, and the second is our generalization of the Weyl formula
for the characters. Remarkably, for all groups and representations this quantity can assume
only three values: 0, -1 and +1. In the adjoint representation of all groups it is always -1.
However, the true value of the average of the Polyakov loop in the adjoint representation is
further reduced (to zero?) after averaging over the fluctuations about the minimum.

This work is a generalization of our previous work on $SU(N)$~\cite{Diakonov:2007nv} and $G(2)$~\cite{Diakonov:2008rx,Diakonov:2009jq}
gauge groups.

\section{Dyons in arbitrary gauge group}

One can divide the full partition function \ur{eff-act} into sectors with given eigenvalues of the Polyakov loop
(or $\mbox{\boldmath$\phi$}$), and integrate over the $\mbox{\boldmath$\phi$}$-sectors at the end. In a sector
with given $\mbox{\boldmath$\phi$}$ there are saddle points in the path integral that are
Bogomolny--Prasad--Sommerfield (BPS) monopoles~\cite{BPS}, or dyons. They are (anti) self-dual solutions
of the nonlinear Maxwell equations, $D_\mu^{ab}F^b_{\mu\nu}=0$, with the condition that at spacial infinity
the $A_4$ component of the solution tends to a given matrix \ur{A4}. This is for the gauge where $A_4$ is static
and diagonal. In a gauge-invariant formulation, dyon solutions correspond to fixed eigenvalues of the Polyakov
line \ur{Pol1} at spatial infinity.

For an arbitrary gauge group dyon solutions have been constructed in Ref.~\cite{Davies:2000nw}. There are $r+1$ fundamental
dyons and $r+1$ fundamental anti-dyons in a group of rank $r$. Each sort or kind of dyons is in fact an $SU(2)$
object associated with one of the simple roots of the group. In the static gauge their fields are static.
The last, $(r\!+\!1)$'th dyon is sometimes called the Kaluza--Klein (KK) monopole; its field is time-dependent
but the action density is static, too. It is built on the $SU(2)$ subgroup spanned by the {\em maximally negative
root} $\mbox{\boldmath$\alpha$}_0$. It plays a key role in constructing Kac--Moody algebras applied in string theory,
and also here, in the monopole business. The corresponding roots used to build dyons are illustrated
for the $G(2)$ group in Fig.~1.

\begin{figure}[h]
\includegraphics[width=0.30\textwidth]{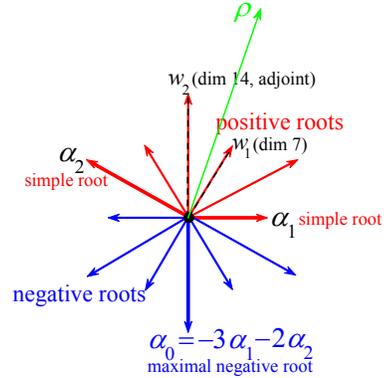}
\caption{Root diagram for the $G(2)$ group. There are three kinds of
dyons, based on simple roots $\alpha_{1,2}$ and the maximal negative root $\alpha_0$.
Also shown is the Weyl vector $\rho$ that gives the preferred direction
of $A_4$ in the confinement phase, and the eldest weights of the two representations
of dimension 7 and 14.}
\end{figure}

The actions of $r+1$ dyons depend on $\mbox{\boldmath$\phi$}$:
\bea\la{actions}
S_i(\mbox{\boldmath$\phi$})&=&\frac{2\pi}{\alpha_s}\nu_i(\mbox{\boldmath$\phi$}),\quad i=0,...,r,\\
\nn
\nu_0(\mbox{\boldmath$\phi$})&=&
\left(1+(\mbox{\boldmath$\phi$}\cdot\mbox{\boldmath$\alpha$}^*_0)\frac{\mbox{\boldmath$\alpha$}_{\rm max}^2}{2}\right),\\
\nn
\nu_i(\mbox{\boldmath$\phi$})&=&
(\mbox{\boldmath$\phi$}\cdot\mbox{\boldmath$\alpha$}^*_i)\frac{\mbox{\boldmath$\alpha$}_{\rm max}^2}{2},
\quad i=1,...,r,
\eea
where $\mbox{\boldmath$\alpha$}^*=2\mbox{\boldmath$\alpha$}/(\mbox{\boldmath$\alpha$}\cdot \mbox{\boldmath$\alpha$})$
are dual roots (coinciding with roots for $A,D,E$ groups). Dyons have a non-linear core of the size
$r_i\sim (\mbox{\boldmath$\alpha$}_i\cdot \mbox{\boldmath$\phi$})/2\pi T$; beyond the core they
have Abelian chromo-electric and -magnetic fields
\begin{figure}[t]
\includegraphics[width=0.32\textwidth]{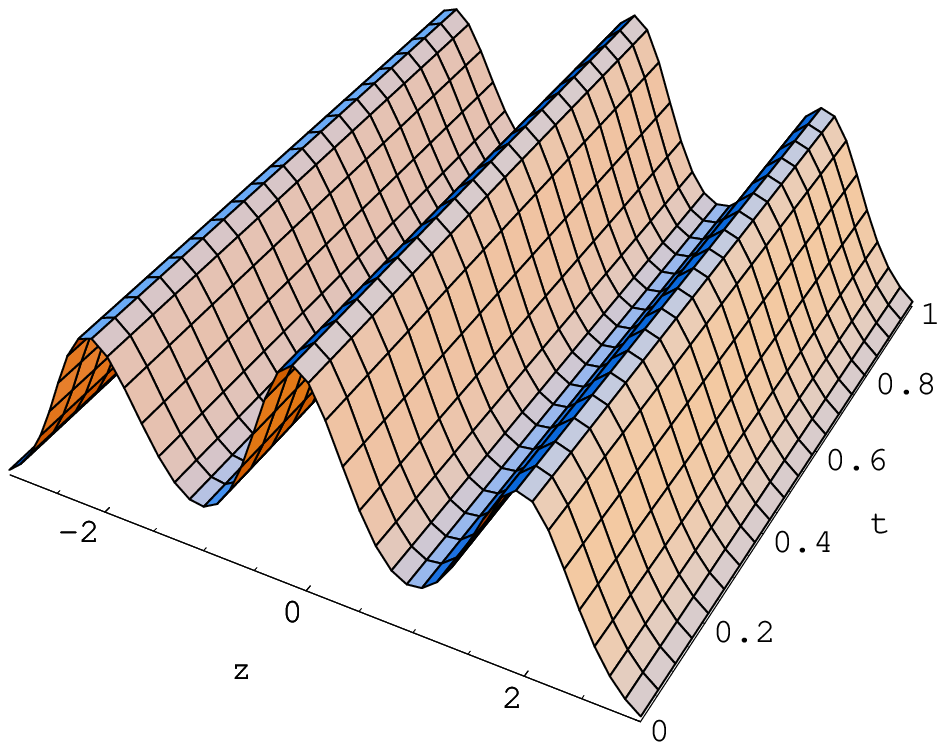}
\includegraphics[width=0.32\textwidth]{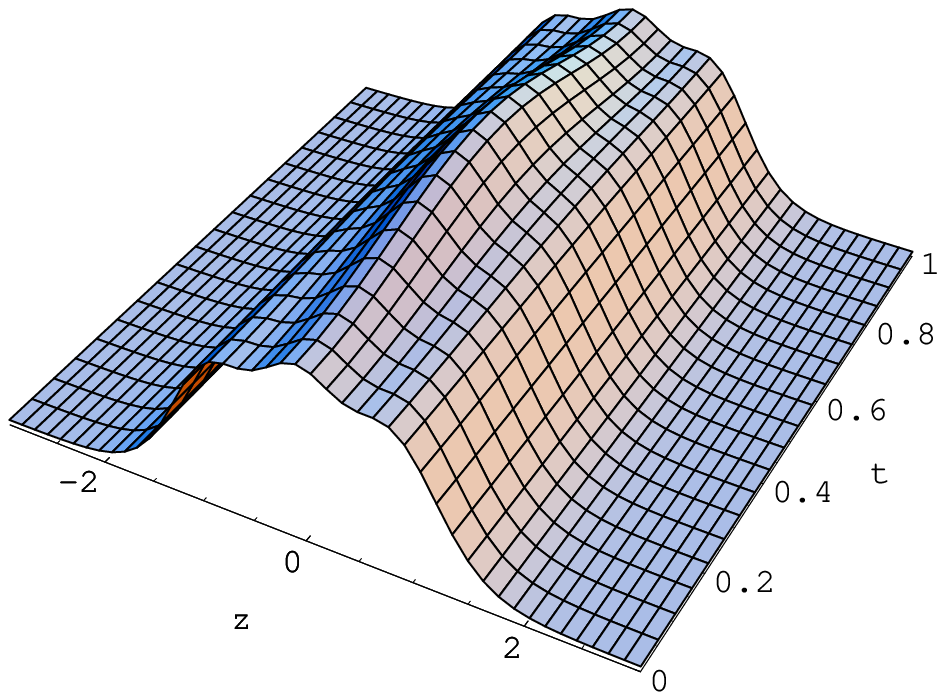}
\includegraphics[width=0.32\textwidth]{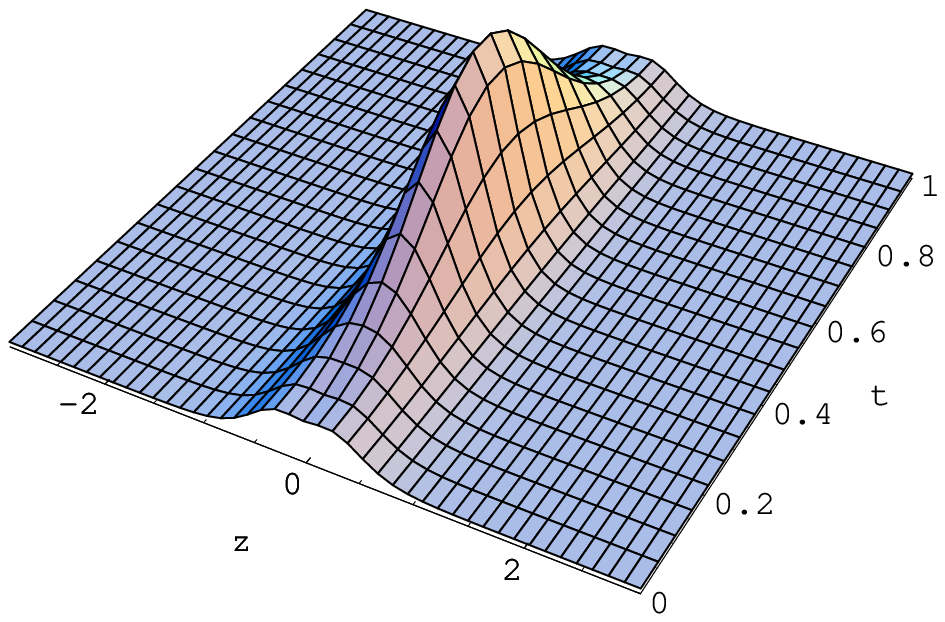}
\caption{Action density inside the $SU(3)$ KvBLL instanton as function of time and
one space coordinate, for large ({\it left}), intermediate ({\it middle}) and small
({\it right}) separations between the three constituent dyons. The full action is the same
for all plots.}
\end{figure}
\beq
\pm B_r^{(i)}=E_r^{(i)}=\frac{1}{2r^2}\,(\mbox{\boldmath$\alpha$}^*_i\cdot {\bf H}),\quad i=0,1,\ldots,r.
\la{BE}\eeq

Our aim is to evaluate the YM partition function in a sector with given $\mbox{\boldmath$\phi$}$ semiclassically,
{\it i.e.} saturating it by the dyon saddle points. The resulting free energy will be a non-trivial function
of $\mbox{\boldmath$\phi$}$.

According to the general semiclassical logic, one has to compute the small-oscillation determinant about the saddle
points. For individual dyons, however, such determinant is infrared divergent because of the Coulomb asymptotics
of the background field. It means that the statistical weight of individual dyons is zero (although the classical
action is finite). However, one can group $r+1$ kinds of dyons into electric- and magnetic-neutral clusters such
that the small-oscillation determinant about the cluster is finite. As one arranges such a neutral cluster, it is
important that it still remains a saddle point, {\it i.e.} satisfies the equation of motion with a given asymptotic
value of $A_4$.

Fortunately, such solutions exist and are known analytically; they are called calorons or instantons
with nontrivial holonomy~\cite{Kraan:1998kp:1998kp,Lee:1997vp}. We call them KvBLL instantons for short. The separations of constituent
dyons centers are collective coordinates of the solution, and form its moduli (parameter) space. For large
separations the KvBLL instanton action density consists of $r+1$ isolated static action densities of individual
dyons, see Fig.~2, left. If we move the constituents closer, the solution tells us that the action density becomes
time-dependent. This is because one of the constituents, the KK monopole, has a time-dependent core, even in a static
gauge. In the limit of complete merger of constituent dyons, the KvBLL instanton becomes a $4d$ lump, a time-dependent
`process'. At low temperatures or $\mbox{\boldmath$\phi$}\to 0$ it becomes the standard BPST instanton
which is $O(4)$ symmetric.

Let us determine how many constituent dyons enter the KvBLL instanton, to make it a neutral cluster. As seen from
\Eq{BE} the neutrality condition is that the sum of all dual roots (including the maximal negative one) is zero.
Since the dual of the maximal negative root $\mbox{\boldmath$\alpha$}^*_0$ is not linearly independent as an $r$-dimensional
vector it can be expanded in the other dual roots, $\mbox{\boldmath$\alpha$}^*_0=-\sum_{i=1}^rk_i^*\mbox{\boldmath$\alpha$}^*_i$,
or
\beq
\sum_{i=0}^rk_i^*\mbox{\boldmath$\alpha$}^*_i=0,
\la{identity}\eeq
where the integers $k_i^*$ are called dual Kac labels, or co-marks; by construction $k_0^*=1$. The sum
\beq
c_2=\sum_{i=0}^rk_i^*
\la{dual_Coxeter}\eeq
is called the dual Coxeter number.

\Eq{identity} tells us that in order to build an
electric- and magnetic-neutral object (the KvBLL instanton with a nontrivial holonomy) one needs $c_2$
fundamental dyons, some of which may have multiplicities other than unity. For example, in the $SU(N)$
group all dual Kac labels are unity, such that instantons consists of $N$ dyons, one for each kind.
In the $G(2)$ gauge theory the three dual Kac labels are $k^*=(1,1,2),\;c_2=4$, meaning
that in this case the KvBLL instanton is made of four dyons but only of three kinds: one kind is taken twice,
namely the one built from $\mbox{\boldmath$\alpha$}^*_2$.
As a curiosity, the dual Kac labels for the exceptional $E(8)$ group are $k^*=(1,2,3,4,6,5,4,3,2),\;c_2=30$,
meaning that the KvBLL instanton of the $E(8)$ gauge theory is built out of 30 dyons of 9 kinds. Other Lie
groups can be easily studied using Ref.~\cite{Bourbaki}.

The one-loop small-oscillation determinant about the neutral KvBLL instanton is IR finite, moreover it has been
computed {\em exactly} for the $SU(2)$ case, and its main features found out for a general $SU(N)$~\cite{Diakonov:2004jn}.
The determinant gives the perturbative potential energy \ur{P-pert} multiplied by the 3-volume, since
in most of the space outside the cores there is just a constant $A_4=2\pi T (\mbox{\boldmath$\phi$}\cdot {\bf H})$,
and other factors.

The weight with which a neutral cluster of dyons appears in the partition function is proportional to the
Jacobian made of zero modes, that depends on the separation of dyons; it has been computed exactly for the
$SU(N)$ group in Refs.~\cite{Kraan:1998pn,Diakonov:2005qa}, but is easily generalizable to arbitrary group. This Jacobian
is the determinant of an $(r\!+\!1)\times(r\!+\!1)$ matrix built of Coulomb bonds, with dyons actions at
the diagonal:
\bea\nn
G_{ij}&=&\delta_{ij}\left(4\pi\nu_i(\mbox{\boldmath$\phi$})
-\sum_{k\neq i}^r\frac{(\mbox{\boldmath$\alpha$}^*_i\cdot\mbox{\boldmath$\alpha$}^*_k)}{r_{ik}}
\frac{\mbox{\boldmath$\alpha$}_{\rm max}^2}{2}\right)\\
&+&\left.\frac{(\mbox{\boldmath$\alpha$}^*_i\cdot\mbox{\boldmath$\alpha$}^*_j)}{r_{ij}}
\frac{\mbox{\boldmath$\alpha$}_{\rm max}^2}{2}\right|_{i\neq j}.
\la{G1}\eea
It leads to the attraction of dyons whose roots are not orthogonal.
A very non-trivial fact is that this Jacobian remains exact at all separations between dyons, even when
their cores overlap. In fact, it is the main factor that determines the interaction between dyons.

\section{Ensemble of many dyons of different kind}

In Ref.~\cite{Diakonov:2007nv} (dealing with $SU(N)$) a straightforward generalization of \ur{G1} was promoted
to the ensemble of any number of dyons. The resulting statistical mechanics was shown to be identical
to a $3d$ quantum field theory involving $r+1$ `electric' fields $\mbox{\boldmath$\phi$}({\bf x})$,
$r+1$ dual (`magnetic') fields, and $r+1$ complex ghost (anticommuting) fields. This quantum field theory appears to be exactly
solvable owing to exact cancelation between boson and ghost loops. Although the theory has nontrivial correlation
functions (in particular leading to the Debye screening of dyons, and to the area behavior of Wilson loops~\cite{Diakonov:2007nv,Diakonov:2009jq}),
the free energy itself is remarkably simple: it is the same as if there were no Coulomb interactions between dyons at all.

Therefore, as far as the free energy is concerned, it is sufficient to evaluate the partition function of
the ensemble of dyons by keeping only the product of constants $4\pi\nu_i(\mbox{\boldmath$\phi$})$
(being in fact nothing but the actions of individual dyons, see \Eq{actions}) on the diagonal of \ur{G1}:
\beq
Z^{\rm dyons}=\sum_{K}\frac{(4\pi f V \nu_0)^{k^*_0 K}\ldots (4\pi f V \nu_r)^{k^*_rK}}{(k^*_0K)!\ldots (k^*_rK)!},
\la{Z2}\eeq
where $K$ is the number of instantons, {\it i.e.} the number of neutral dyon clusters made of $c_2=k^*_0+\ldots + k^*_r$
dyons, $V$ is the $3d$ volume and $f$ is the `fugacity'. With one loop renormalization performed
when the small-oscillation determinant is computed~\cite{Diakonov:2004jn}, it is expressed through the scale parameter
$\Lambda$ (in the Pauli--Villars scheme), and the 't Hooft coupling $\lambda$ which, for arbitrary group,
we define as $\lambda\equiv c_2\alpha_s/(2\pi)$:
\beq
f\approx\frac{c_2^2}{16\pi^3\lambda^2}\frac{\Lambda^4}{T}\prod_{i=0}^r\left(\frac{\alpha_{\rm max}^2}{\alpha_i^2}\right)^{\frac{k_i^*}{c_2}}.
\la{f1}\eeq
At one loop level $\lambda$ in \Eq{f1} is not renormalized but eventually it will become a running coupling constant
whose argument is expected to be of the order of the equilibrium density of dyons. There are $k_i^*K$ dyons of
kind $i$ in the ensemble. The factorials in \Eq{Z2} are introduced such that identical configurations of dyons
are not counted more than once.

The factor $(4\pi f V)$ is dimensionless and much larger than unity at large volumes, and one can sum over
$K$ in \Eq{Z2} by the saddle-point method. Using the Stirling asymptotics
for the factorials we obtain the free energy of the ensemble, as function of
$\mbox{\boldmath$\phi$}$:
\bea\la{F}
F^{\rm nonpert}(\mbox{\boldmath$\phi$})\!\!\!\!&=&\!\!\!\!-T\ln\; Z^{\rm dyons}\\
\n
&=&-4\pi f V T c_2\prod_{i=0}^r\left(\frac{\nu_i(\mbox{\boldmath$\phi$})}{k^*_i}\right)^{k^*_i/c_2}\!\!\!\!\!,\\
\la{sum-rule}
k^*_0\nu_0(\mbox{\boldmath$\phi$})+\!\!&\ldots&\!\! + k^*_r\nu_r(\mbox{\boldmath$\phi$})=1.
\eea
The last relation follows from \Eqs{actions}{identity}: the sum of dyons actions inside a neutral
cluster is equal to the standard instanton action~\footnote{Actually the overall coefficient in \ur{F}
has to be doubled as due to anti-dyons~\cite{Diakonov:2007nv,Diakonov:2009jq}.}.

We have now to find the minimum of the dyon-induced free energy \ur{F} as function of $\mbox{\boldmath$\phi$}$.
We write down the constraint \ur{sum-rule}
via a Lagrange multiplier and then minimize \ur{F} in all $r\!+\!1$ $\nu$'s independently. The result
of the minimization is that all $\nu$'s are equal,
\beq
\nu_0=\ldots =\nu_r =\frac{1}{c_2},
\la{nu-opt}\eeq
meaning that the minimum corresponds to all dyon actions \ur{actions} being equal. Finally, we make use of the
fact from Lie algebra that $(\mbox{\boldmath$\rho$}\cdot \mbox{\boldmath$\alpha$}^*_i)=1$ for all $i=1,\ldots,r$,
where $\mbox{\boldmath$\rho$}$ is the half-sum of positive roots, the Weyl vector. Recalling the definition
of $\nu$'s as the dyon actions \ur{actions}, we obtain the optimal value of $\mbox{\boldmath$\phi$}$:
\beq
\mbox{\boldmath$\phi$}_{\rm min}={\bf v}={\bf\rho}\;\frac{2}{c_2\mbox{\boldmath$\alpha$}_{\rm max}^2},
\la{v-opt-1}\eeq
that is proportional to the Weyl vector
. At the minimum,
\beq
F^{\rm nonpert}({\bf v})=-4\pi f T V\prod_{i=0}^r (k^*_i)^{-\frac{k^*_i}{c_2}}= {\cal O}(c_2^2\Lambda^4).
\la{F-min}\eeq
Note that the free energy is temperature-independent, and can be used all the way to $T=0$.
\begin{figure}[h]
\begin{picture}(400,240)
\put(-35,122)
{
\includegraphics[width=0.24\textwidth]{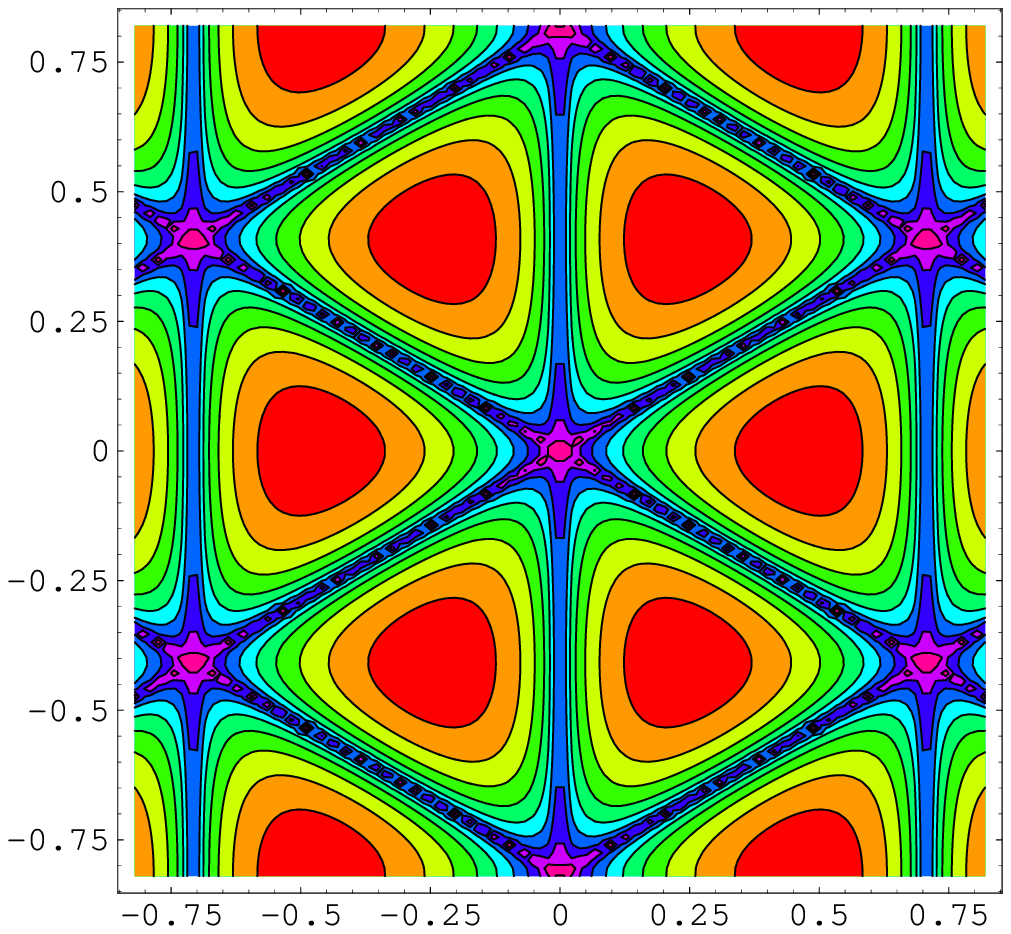}
\includegraphics[width=0.24\textwidth]{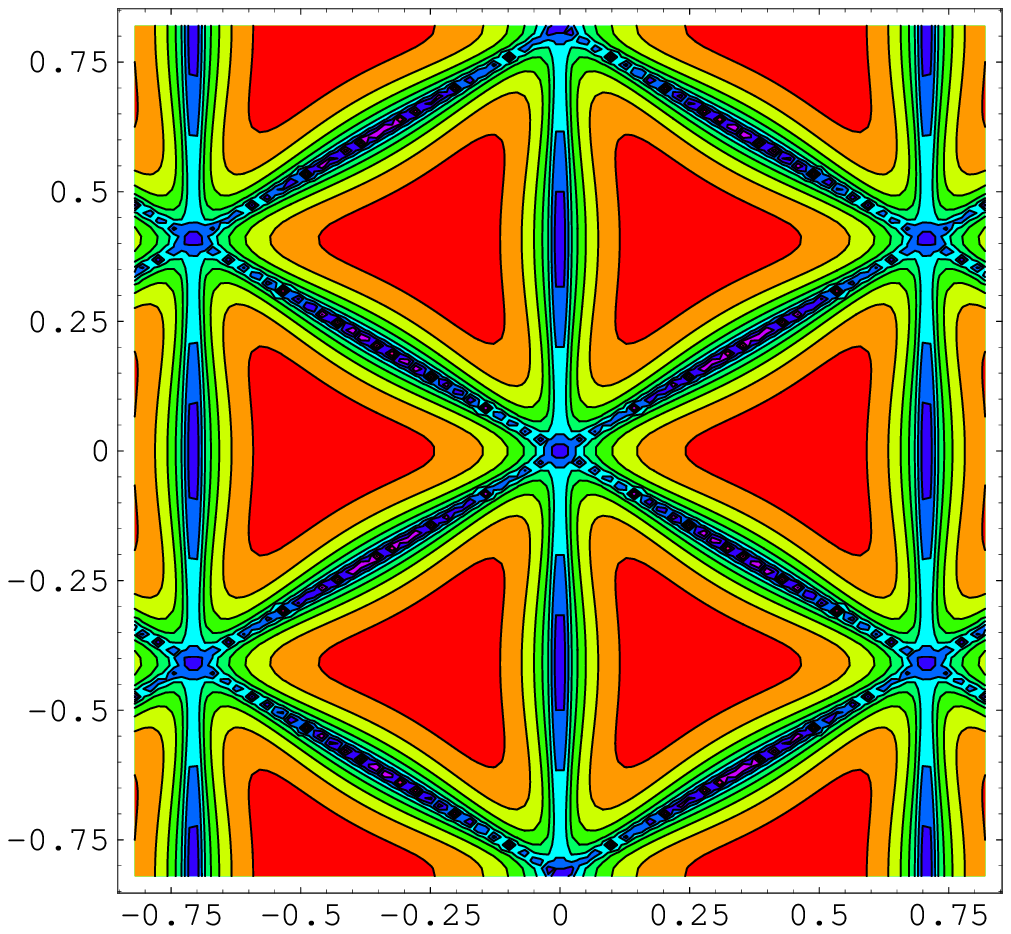}
\includegraphics[width=0.24\textwidth]{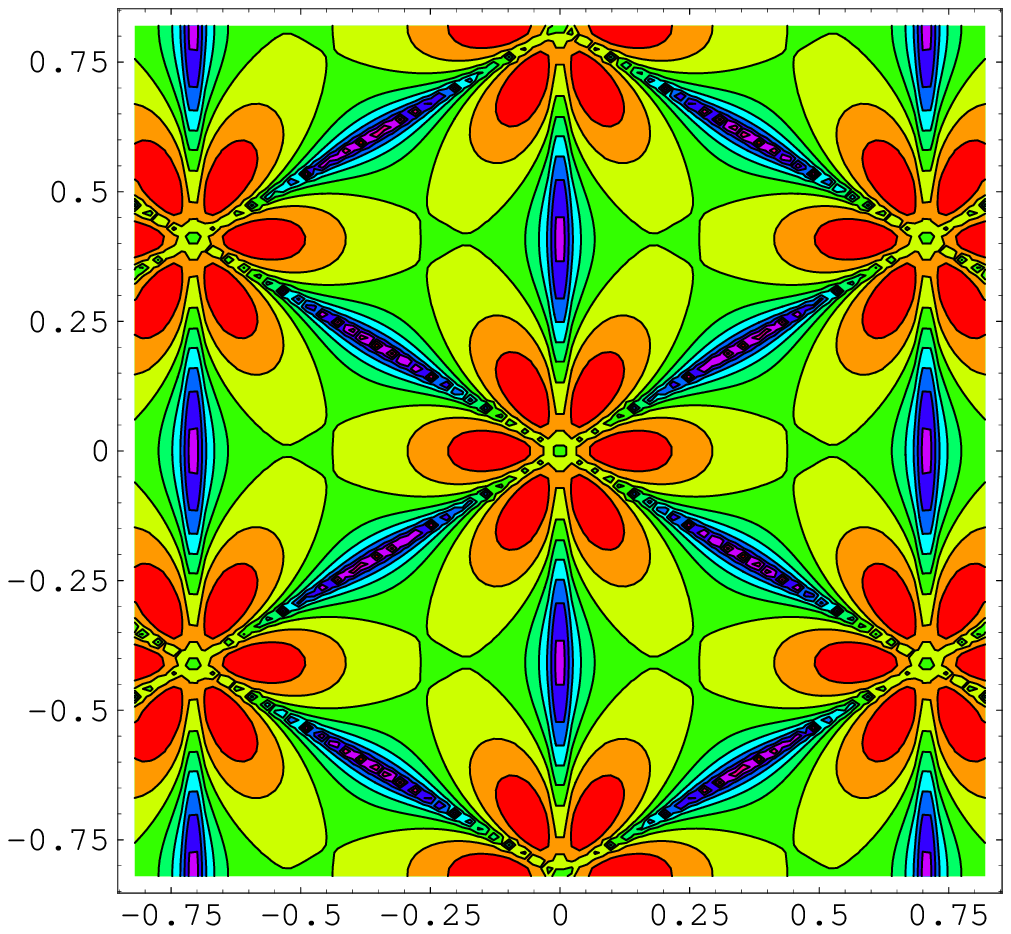}
\includegraphics[width=0.24\textwidth]{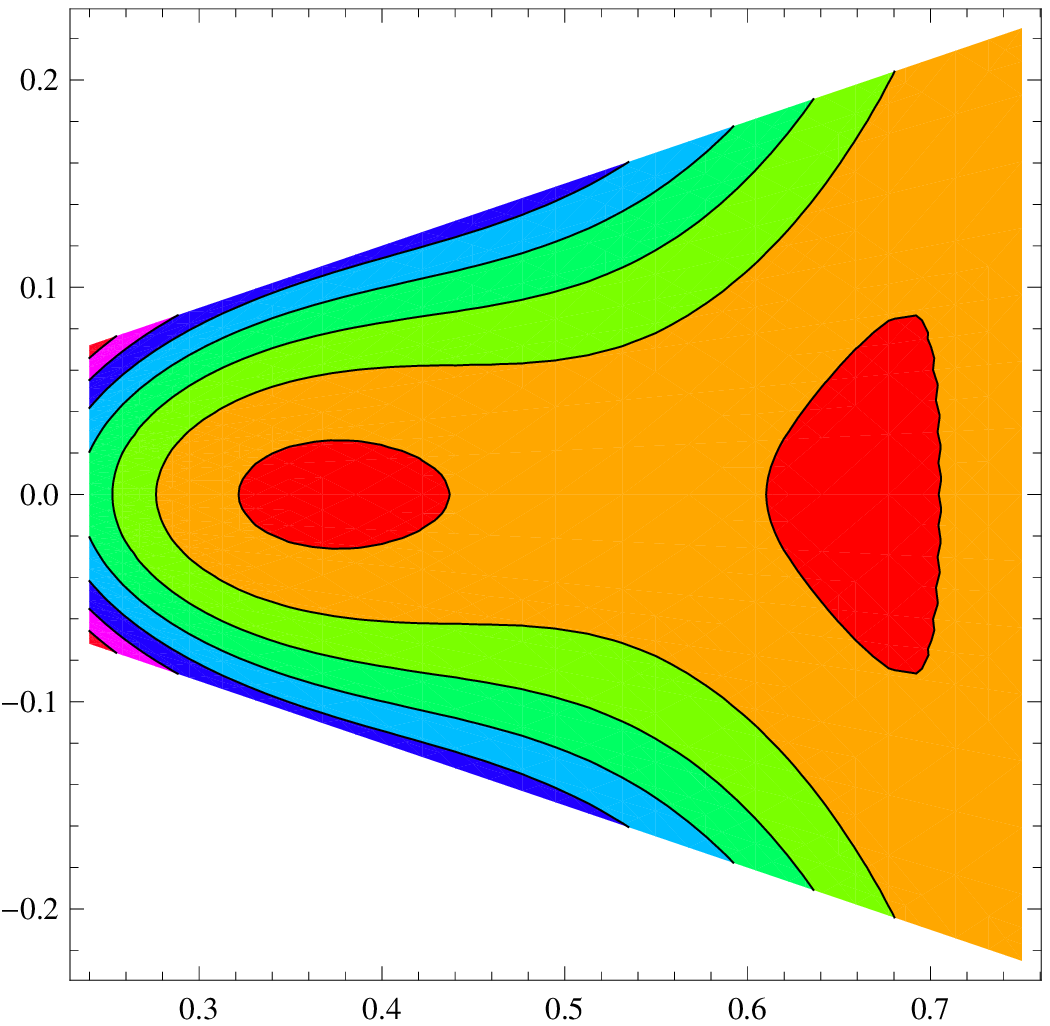}
}
\put(-35,2)
{
\includegraphics[width=0.24\textwidth]{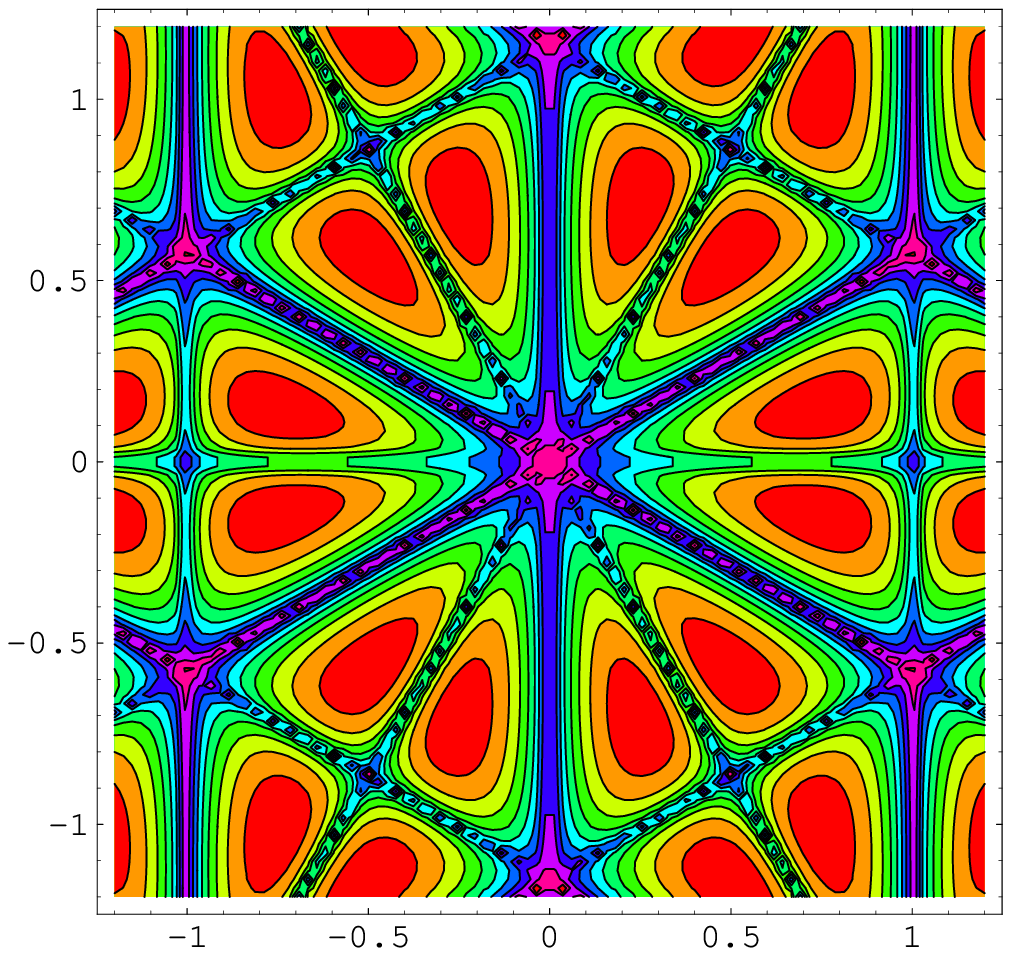}
\includegraphics[width=0.24\textwidth]{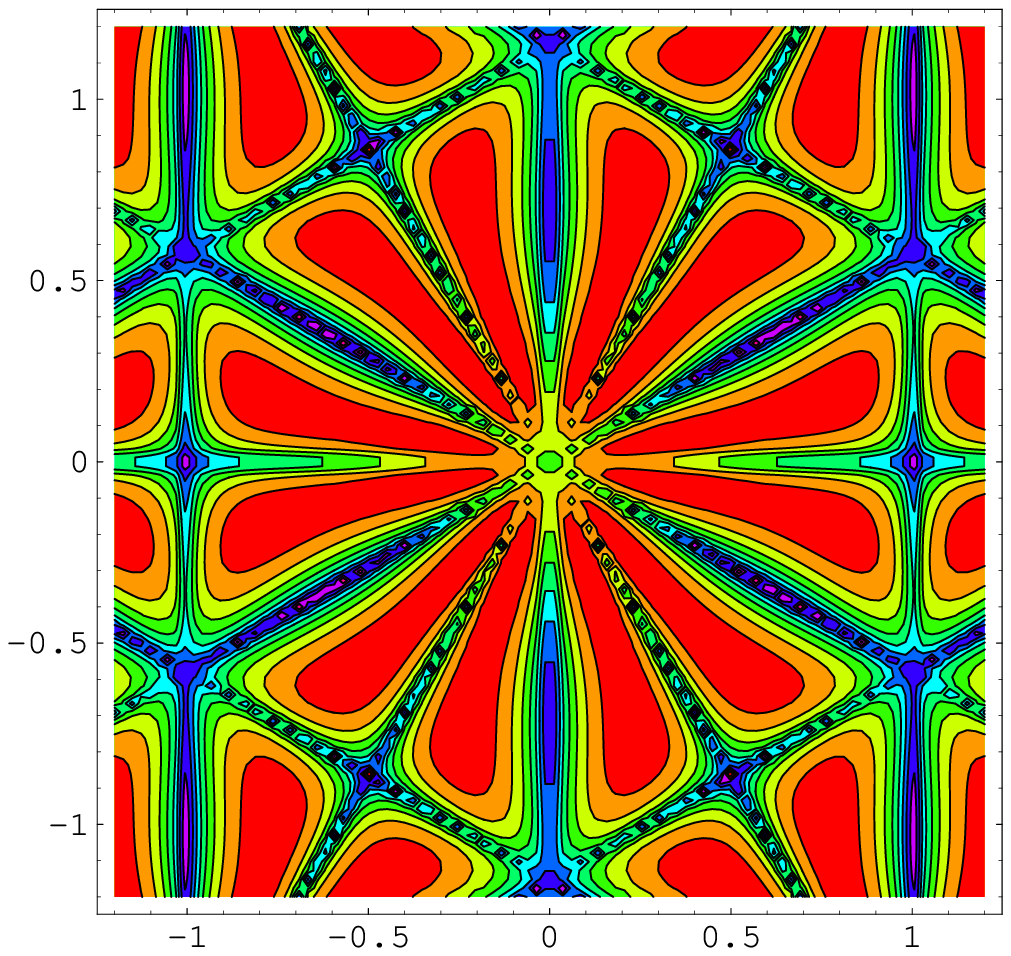}
\includegraphics[width=0.24\textwidth]{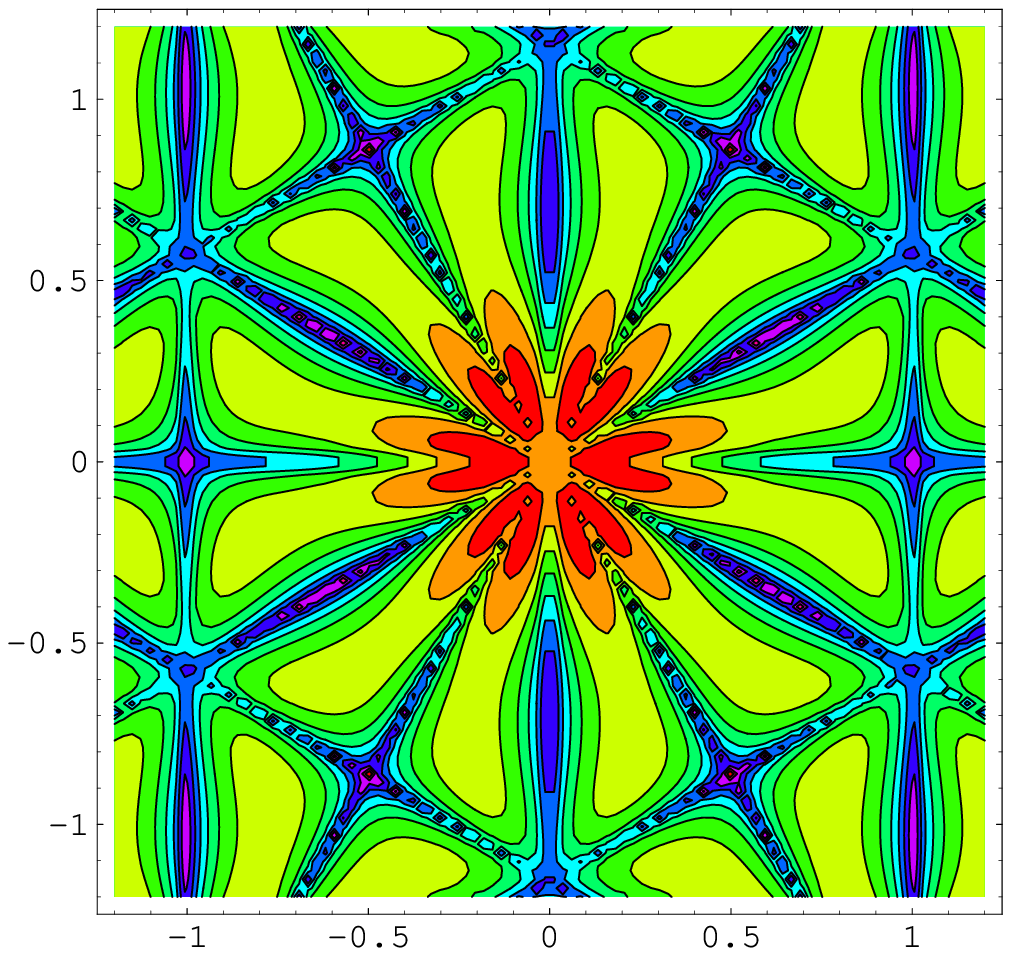}
\includegraphics[width=0.24\textwidth]{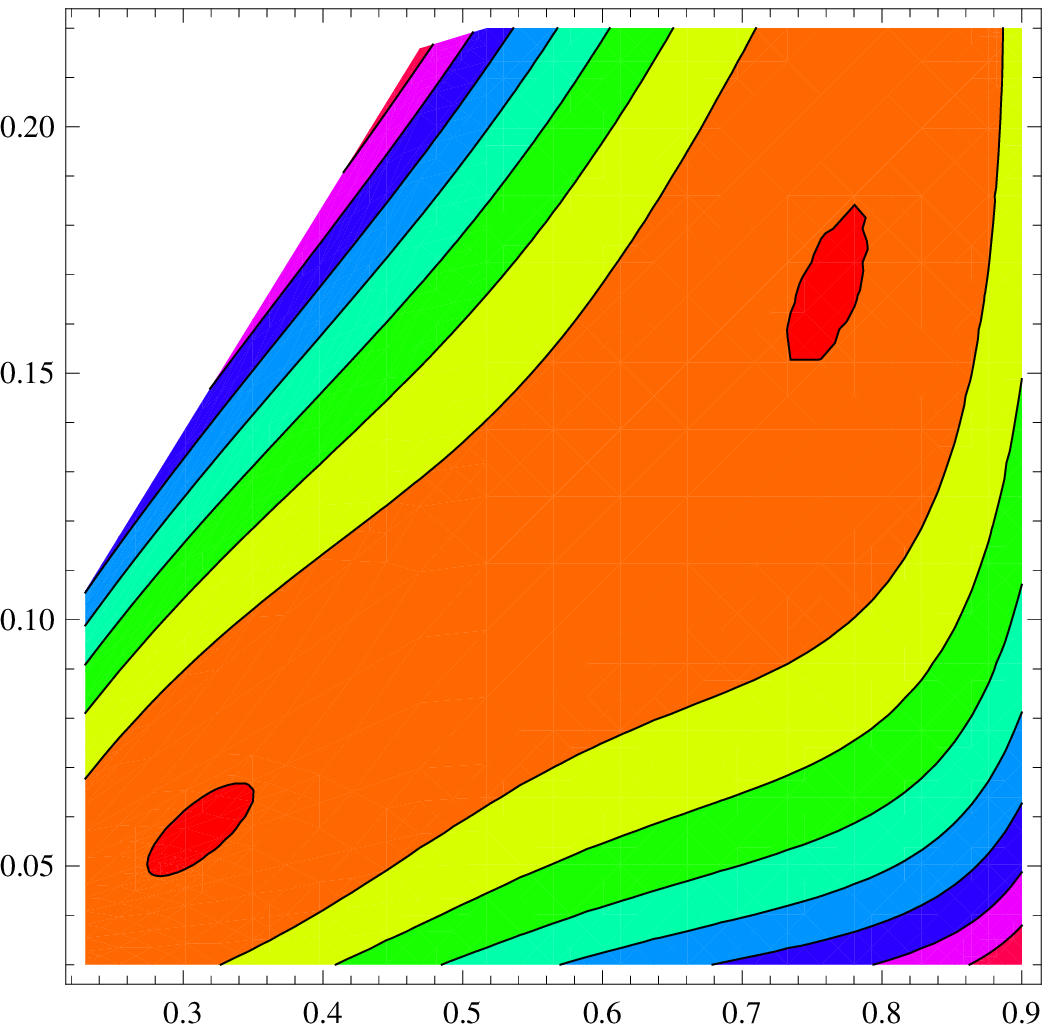}
}
\end{picture}
\caption{Contour plots for the potential energy (perturbative plus nonperturbative) in the
$\mbox{\boldmath$\phi$}=(\phi_1,\phi_2)$ plane for $SU(3)$ ({\it upper row}) and $G(2)$ ({\it lower row})
gauge groups. The first column is at zero temperatures, the second column is at critical temperatures,
the third column is at high temperature. The last column is a zoom at critical temperature, which
clearly exhibits a coexistence of two minima with different values of the Polyakov line,
characteristic of the 1st order transition.}
\end{figure}

The logic, thus, is as follows: One assumes that $A_4$
in the vacuum
is, on the average,
nontrivial, then dyons emerge as saddle points in the partition function. The semiclassical evaluation
of the free energy induced by dyons gives the minimum at a nontrivial $A_4$, thus justifying
its introduction in the first place. It is remarkable that the `optimal' vacuum value
of $A_4$ is given by a universal relation \ur{v-opt-1} valid for any gauge group.

\section{Confinement}

The first thing one has to check is the average of the Polyakov line: at the minimum \ur{v-opt-1}
it is given by \Eq{Pol2}. Taking, {\it e.g.} the $SU(N)$ gauge group we obtain that $\Tr L=0$
in the fundamental (quark or antiquark) representation, and also in any $N$-ality nonzero representation.

An interesting case is the $G(2)$ gauge group studied recently by several lattice groups~\cite{G2}.
The Weyl vector is~\cite{Bourbaki} $\mbox{\boldmath$\rho$}=(-1,-2,3)$ (it is convenient to present the $2d$
vectors of $G(2)$ as belonging to the hyperplane in $3d$ with the zero sum of 3 components), $\alpha_{\rm max}^2=6,\;
c_2=4$, so that the optimal ${\bf v}=(1/12)(-1,-2,3)$. The eldest weight of the lowest $7d$ representation
is ${\bf W}=(0,-1,1)$, the other weights are obtained by permutations. One finds from the first line in \Eq{Pol2}
that the seven preferred eigenphases of the Polyakov loop are
\beq
G(2):\qquad 0,\;\pm \frac{\pi}{6},\;\pm\frac{4\pi}{6},\;\pm\frac{5\pi}{6},
\la{G2-phases}\eeq
which sum into $\Tr L=0$. A quicker way to get this is to use the second line in \Eq{Pol2}. For a more detailed
analysis of $G(2)$ see Ref.~\cite{Diakonov:2009jq}.

By the same argument, the vector and spinor representations of the $SO(2n),\;SO(2n+1)$ and $Sp(n)$ groups,
as well as the lowest-dimension representations of the exceptional $F(4),\;E(6),\;E(7)$ groups -- all have
$\Tr L=0$ at the minimum, regardless of what is the group center.

A more subtle probe is the correlation function of Polyakov lines at large separations, and large spatial Wilson loops,
in various representations. We have shown in Ref.~\cite{Diakonov:2007nv} for the $SU(N)$ group that asymptotically
the linear rising potential exists only for nontrivial $N$-ality representations, with the asymptotic
string tension proportional to $\sin(\pi k/N)$ (the ``$k$-strings''). At intermediate separations between heavy
probes the effective potential can be anything. The mechanism of the exponential decay of the Polyakov lines
correlations is the appearance of the Debye screening of the dual potential in the `plasma' of dyons.

Calculation of large spatial Wilson loops yields {\em exactly the same string tensions} at small temperatures,
as it should be. It indicates that our theory ``knows'' about the restoration of Euclidean $O(4)$ symmetry
in the limit of zero temperatures, despite its explicit $3d$ formulation. It looks the more striking
that the mechanism of the area law is the appearance of a double-layer soliton for the dual potential,
pinned to the surface spanning the loop~\cite{Diakonov:2007nv}. It looks physically quite different from the
mechanism of the electric string generation.

For a general group, all possible asymptotic string tensions squared for confining representations are
eigenvalues of the $r\times r$ Debye matrix (squared),
\beq
{\cal M}^2_{mn}=\frac{8\pi f}{T}\prod_{i=0}^r(k^*_i)^{-k^*_i/c_2}
\sum_{i=0}^rk^*_i \mbox{\boldmath$\alpha$}^*_m(i)\mbox{\boldmath$\alpha$}^*_n(i).
\la{M}\eeq

\section{Deconfinement}

The dyon-induced nonperturbative potential energy \ur{F} ``wants'' $\mbox{\boldmath$\phi$}$ (or $A_4$) to be proportional
to the Weyl vector, while the perturbative potential \ur{P-pert} ``wants'' it to be zero or, if the group has a nontrivial
center, corresponding to any element of the center.

The perturbative potential energy scales as $T^4$ with respect to the nonperturbative one, therefore at low $T$
it is irrelevant, so the system is in the confinement phase. At large $T$ it prevails. At a critical $T_c$
computable from the string tension (and both of them from $\Lambda$) the system undergoes the deconfinement
transition. It is of the second order for $SU(2)$ and of the first order for all gauge groups we have studied
so far, including $SU(N),\;N>3,$ and the centerless $G(2)$~\cite{Diakonov:2007nv,Diakonov:2009jq}, see Fig.~3. This is in accordance
with lattice findings~\cite{G2}. The presence or the absence of a nontrivial center of the group has no relevance
to the fact of the phase transition, and its nature~\cite{Holland:2003kg}.

An important manifestation of confinement is the absence of massless gluons in the spectrum, in particular
the perturbative Stefan--Boltzmann contribution of gluons to the free energy,
$-(\pi^2T^4/45)\times({\rm group\;dimension})$,
must be absent. In the confinement phase it is, indeed, canceled with a high precision by the perturbative
potential energy \ur{P-pert} computed at the optimal $\mbox{\boldmath$\phi$}_{\rm min}\!=\!{\bf v}$ \ur{v-opt-1}.
For example, in $SU(N)$ the perturbative potential energy at the optimal (confining) point is
$(\pi^2T^4\!\!/\!45)(N^4\!-\!1)/N^2$~\cite{Diakonov:2008rx} whereas the Stefan--Boltzmann law gives $-(\pi^2T^4/45)(N^2\!-\!1)$.
The leading ${\cal O}(N^2)$ piece in the $T$-dependence of free energy is canceled! [We wonder how else
can it be canceled if not by forcing $A_4$ to be the Weyl vector.]
For $E(8)$, the group dimension is 248, whereas the perturbative potential
at the optimal point \ur{v-opt-1} gives 248.84 with the opposite sign. The cancelation is not exact, however
there are unaccounted quantum corrections to the free energy. There is a good chance that all ${\cal O}(T^4)$ terms
in the free energy will be absent since there are no massless degrees of freedom left in the theory.
Actually the spectrum is determined by string excitations but we do not consider them here.

{\em To summarize}: dyons exist as saddle points in the YM partition function if $A_4$ or, more precisely,
the eigenvalues of the Polyakov line, are nontrivial. Being admitted, dyons induce a nonperturbative potential
energy as function of $A_4$, that has a minimum at a universal value related to the Weyl vector of the gauge group,
making the semiclassical considerations self-consistent. Deconfinement phase transition happens as
a result of the competition between perturbative and dyon-induced energies as function of $A_4$, irrespectively
of whether the gauge group has a nontrivial center, or not. In the confinement phase, dyons form a kind of
multi-component plasma; the appearance of the Debye screening mass is, physically, the reason for the area
law for large Wilson loops and for the asymptotic linear rising potential for probe sources, where one expects it.
On the whole, dyons produce an appealing picture of confinement and deconfinement, for any gauge group.

\begin{theacknowledgments}
This work has been supported in part by Russian Government grants RFBR-09-02-01198 and RSGSS-65751.2010.2.
We also acknowledge support by the DFG at Bochum University, Germany. D.D. is grateful to the organizers
of the Quark Confinement conference for partially supporting his stay in Madrid, and for a creative atmosphere
at the meeting.
\end{theacknowledgments}

\bibliographystyle{aipproc}   

\begin{thebibliography}{99}

\bibitem{Gross:1980br}
D.J.~Gross, R.D.~Pisarski and L.G.~Yaffe, Rev.\ Mod.\ Phys. {\bf 53}, 43 (1981)

\bibitem{Diakonov:2004kc}
D.~Diakonov and M.~Oswald, Phys.\ Rev.\ D {\bf 70}, 105016 (2004)

\bibitem{Weiss:1980rj}
N.~Weiss, Phys.\ Rev.\ D {\bf 24}, 475 (1981);
Phys.\ Rev.\ D {\bf 25}, 2667 (1982)

\bibitem{Diakonov:2007nv}
D.~Diakonov and V.~Petrov, Phys.\ Rev.\ D {\bf 76}, 056001 (2007)

\bibitem{Diakonov:2008rx}
D.~Diakonov and V.~Petrov, AIP Conf. Proc. {\bf 1134}, 190 (2009)

\bibitem{Diakonov:2009jq}
D.~Diakonov, Nucl. Phys. Proc. Suppl. {\bf 195}, 5 (2009)

\bibitem{Bourbaki}
N.~Bourbaki, {\it Groupes et alg\`ebres de Lie}, Hermann (1968), Table after Ch. VI

\bibitem{BPS}
E.B.~Bogomolny, Sov.\ J.\ Nucl.\ Phys. {\bf 24}, 449 (1976) [Yad.\ Fiz.\ {\bf 24}, 861 (1976)];
M.K.~Prasad and C.M.~Sommerfield, Phys. Rev. Lett. {\bf 35}, 760 (1975)

\bibitem{Davies:2000nw}
N.M.~Davies, T.J.~Hollowood and V.V.~Khoze, J.\ Math.\ Phys.\ {\bf 44}, 3640 (2003)

\bibitem{Kraan:1998kp}
T.C.~Kraan and P.~van Baal, Phys.\ Lett.\ B {\bf 428}, 268 (1998);
Nucl.\ Phys.\ B {\bf 533}, 627 (1998);
Phys. Lett. {\bf B435}, 389 (1998)

\bibitem{Lee:1997vp}
K.~Lee and P.~Yi, Phys.\ Rev.\ D {\bf 56}, 3711 (1997);
K.~Lee and C.~Lu, Phys.\ Rev.\ {\bf 58}, 025011 (1998)

\bibitem{Diakonov:2004jn}
D.~Diakonov, N.~Gromov, V.~Petrov and S.~Slizovskiy, Phys.\ Rev.\ D {\bf 70}, 036003 (2004);
N.~Gromov, in: Proc. NATO Advanced Study Institute and EU Hadron Physics 13 Summer Institute,
St. Andrews, Scotland, 22-29 Aug 2004, p. 411, arXiv:hep-th/0701192;
S.~Slizovkiy, Phys.\ Rev.\ D {\bf 76}, 085019 (2007)

\bibitem{Kraan:1998pn}
T.C.~Kraan, Commun. Math. Phys. {\bf 212}, 503 (2000)

\bibitem{Diakonov:2005qa}
D.~Diakonov and N.~Gromov, Phys. Rev. {\bf D72}, 025003 (2005)

\bibitem{G2}
J.~Greensite, K.~Langfeld, S.~Olejnik, H.~Reinhardt and T.~Tok, Phys.\ Rev.\ D {\bf 75}, 034501 (2007);
M.~Pepe and U.-J.~Wiese, Nucl.\ Phys.\ B {\bf 76}, 21 (2007);
G.~Cossu, M.~D'Elia, A.~Di~Giacomo, B.~Lucini and C.~Pica, JHEP 0710 (2007) 100;
J.~Danzer, C.~Gattringer and A.~Maas, JHEP 0901 (2009) 024

\bibitem{Holland:2003kg}
K.~Holland, M.~Pepe and U.J.~Wiese, Nucl.\ Phys.\ B {\bf 694}, 35 (2004).

\end{thebibliography}

\end{document}